# Spectro-temporal shaping of broadband optical wavepackets via programmable on-chip photonics

**Bruno P. Chaves[1], Jérémy Saucourt[1], Van Thuy Hoang[1,2], Alexis Bougaud[1], Yassin Boussafa[1], Erwan Ferrandon[1], Sai T. Chu[3], Brent E. Little[4], David J. Moss[5], Roberto Morandotti[6], Vincent Couderc[1], Benjamin Wetzel[1*]**

*1. XLIM Research Institute, CNRS UMR 7252, Université de Limoges, Limoges 87060, France*
*2. School of Electrical and Mechanical Engineering, Adelaide University, Adelaide, SA 5005, Australia*
*3. Department of Physics, City University of Hong Kong, Tat Chee Avenue, Kowloon, Hong Kong, China*
*4. QXP Technologies Inc., Xi'an, China*
*5. Optical Sciences Centre, Swinburne University of Technology, Hawthorn, VIC 3122, Australia*
*6. Institut National de la Recherche Scientifique - Centre Énergie Matériaux Télécommunications, Varennes J3X 1S2, Canada*
**e-mail: benjamin.wetzel@xlim.fr*

The ability to precisely shape the properties of light over multiple degrees of freedom constitutes the foundation of modern photonic architectures. Broadband sources with addressable spectro-temporal content are for instance critical for applications spanning biomedical imaging, material processing, lidar, and ultrafast spectroscopy, but also key enabling technologies in quantum information processing, high-capacity communications, and optical computing. Yet, flexible and adjustable spectro-temporal shaping so far remains a significant challenge, as it requires the simultaneous manipulation of both frequency and temporal domains across broad spectral bandwidths and extended timescales, from femtoseconds to nanoseconds. Importantly, conventional approaches to spectro-temporal processing typically rely on bulky and complex optical systems that lack scalability and flexibility, ultimately hampering their utility for applications requiring addressable multiphotonic processes.

Here, we present a framework for spectro-temporal wavepacket control over a broad bandwidth, merging programmable integrated photonics with high-speed optical characterization. Leveraging machine leaning for tailoring nonlinear pulse propagation, we experimentally report scalable and on-demand shaping of broadband ultrafast pulse patterns. We demonstrate the efficacy of our method through reconfigurable power across 400 nm bandwidth with picosecond resolution, with promises for applications where precise control over light is paramount, in particular those needing versatile and controllable multiphoton excitations.

## Introduction

The evolution of photonics has been driven, to a large extent, by the pursuit of the increasingly comprehensive control over light. More recently, two promising trends have emerged in this direction: the development of versatile photonic sources and the quest for multidimensional light control. User-defined control over the properties of optical radiation, particularly after fabrication, can enable more versatile light sources, new photonic functionalities, and systems that are more resilient to environmental and fabrication variations[1]. At the same time, since light is inherently multidimensional, many applications require the simultaneous control of multiple degrees of freedom, motivating recent advances in, for example, spatial control[2–5].

Flexible temporal control, on the other hand, has received comparatively less attention. Many applications, however, rely on both the spectral and temporal profiles of light. Multiphoton imaging provides a prominent example, where different imaging modalities require distinct spectro-temporal profiles to probe different physical or biological properties[6–12]. To illustrate this requirement, Fig. 1a–d shows representative spectro-temporal profiles associated with different imaging modalities. An ideal flexible source for such applications would therefore resemble Fig. 1e, where the spectro-temporal profile can be dynamically programmed according to the requirements of the end user, enabling multimodal imaging within a single broadband and reconfigurable source.

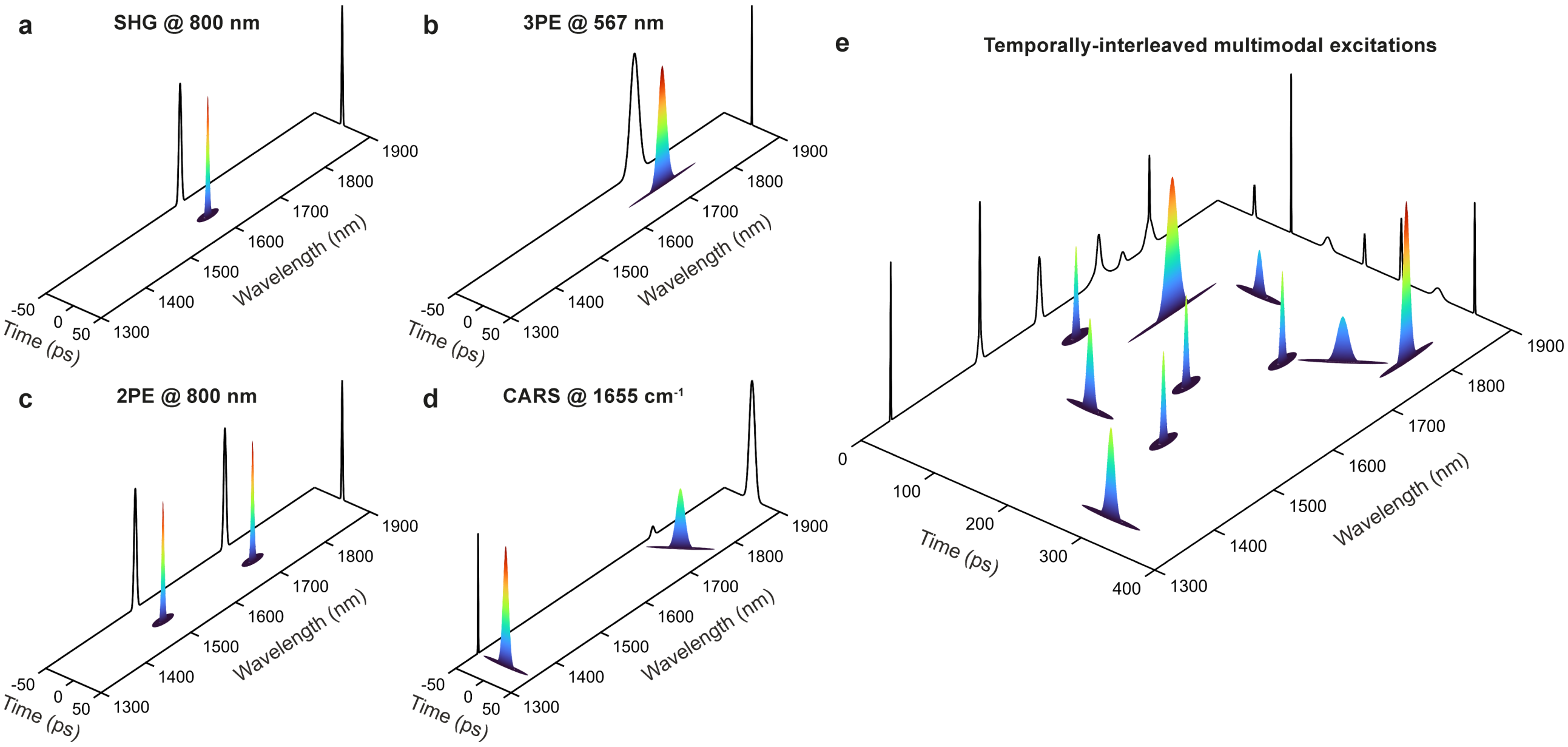

**Figure 1 | Properties of spectro-temporally tailored wavepackets for multiphoton imaging.** Examples of selected spectrograms in the near-infrared region typically required for exciting diverse multiphoton imaging modalities. **a-d** show several wavepackets with spectro-temporal properties desirable for relevant multiphoton excitations. Such illustrative spectrograms include single pulse targeting second-harmonic generation (SHG) at 800 nm (a) or three-photon excitation (3PE) at 567 nm (b), aiming to excite typical fluorescence markers (e.g. *Texas Red*). The spectral bandwidth and associated peak power are directly related to the number of photons involved in the multiphotonic process, to ensure efficient excitation. Dual-pulse spectrograms are also shown for exciting non-degenerate two-photon excitation (2PE) at 800 nm (c), or specific coherent anti-Stokes Raman scattering (CARS) responses (d). Here, we display the example of *Amide I* resonance excitation around 1655 $cm^{-1}$, using in this case a broad excitation with a chirped probe pulse. **e** shows a complex spectrogram, encompassing multiple excitations (including the cases a-d) that are temporally-interleaved, as typically required for multimodal nonlinear imaging via broadband optical wavepackets.

Applications relying on spectro-temporal control often also require broad spectral coverage, i.e. broadband wavepackets. Supercontinuum (SC) generation, which exploits nonlinear propagation to produce extremely broadband optical sources, has become a workhorse for numerous applications, with demonstrated spectral coverage extending from the UV to the mid-IR[13–15]. However, most SC sources remain relatively fixed: their spectro-temporal output cannot be readily reconfigured, and the source is typically designed around a specific application or operating condition. Works on flexible SC sources do exist, mainly exploiting input power adjustment[13,14], pulse shaping[16–19], or waveguide tuning[13,20–22], but these have not demonstrated versatile spectro-temporal control.

Achieving the flexibility required to tailor such broadband nonlinear sources calls for control over the underlying light propagation. Photonic integrated circuits (PICs) offer a promising route towards this goal, combining compactness and scalability with precise control over optical fields[23,24]. In particular, programmable PICs enable the optical propagation and processing of light to be dynamically configured, opening new possibilities for adaptable photonic systems. This has led to them bringing flexibility to applications as diverse as microwave generation, image processing, and telecommunications [1,25–27].

For nonlinear propagation, however, such as that underlying SC generation, the relationship between the input degrees of freedom and the resulting output waveform is highly complex, with no straightforward analytical mapping between the two. Machine-learning (ML) techniques offer a promising approach to navigating this complex relationship, at the cost of requiring large datasets for training[20,28–33]. This becomes particularly challenging for multidimensional optical control, where the acquisition of sufficiently large and information-rich datasets must be performed within a reasonable timeframe.

Here, we address these challenges by combining a programmable PIC with asynchronous optical sampling (ASOPS)[34,35] and ML-driven optimization. Specifically, our approach leverages a programmable delay line (PDL), a class of reconfigurable PICs able to generate adjustable optical pulse trains with sub-picosecond resolution. Over the years, this system was successfully implemented for optical signal processing and optimization[36–39], while numerical works demonstrated further potential for generating user-defined spectro-temporal patterns for applications requiring selective multiphoton excitation[6,32]. On the characterization side, our experimental setup allows for the acquisition the SC spectro-temporal profiles with a measurement time in the scale of just a few seconds. This combination enables the experimental optimization of nonlinear wave propagation using metaheuristic algorithms, with use of an artificial neural network (ANN) further accelerating convergence.

We demonstrate on-demand generation of complex spectro-temporal wavepackets, including user-defined temporal delays at fixed wavelengths, adjustable pulse sequences, synchronized overlaps between multiple wavelengths, and wavelength-specific delays capable of overcoming the dispersion profile of the nonlinear waveguide. Together, these results demonstrate a flexible framework for broadband wavepacket synthesis in which the spectro-temporal output can be dynamically tailored to user-defined requirements rather than being constrained by a fixed source configuration. Such flexibility is particularly relevant to applications including multiphoton microscopy[6–12], material processing[40,41], coherent control[42], quantum signal processing[43–45], and photonic computing[46–49], where precise synchronization and spectral allocation of optical fields are essential.

# Results

## Experimental setup

The experimental setup used for wavepacket spectro-temporal shaping is shown in Fig. 2. An input signal processing stage (red shading), based around an on-chip PDL transforms an initial pulse from a 1550 nm mode-locked laser into a pulse train of femtosecond pulses with a controllable temporal profile, identified by the splitting ratios of several cascaded Mach-Zehnder interferometers (see Methods). After amplification via an Erbium-doped fibre amplifier (EDFA) paired with suitable dispersion management, the tailored pulse pattern is injected into a highly nonlinear fibre (HNLF), seeded in the anomalous dispersion regime, where it undergoes nonlinear fibre propagation (green shading). The individual pulses' broadening dynamics, and their respective interactions, lead to the formation of a broadband output SC featuring a highly structured spectro-temporal content. In our architecture, this output wavepacket can be readily adjusted by dynamically tuning the initial pulse pattern (via reconfigurable on-chip PDL pulse splitting), that ultimately conditions nonlinear fibre pulse propagation.

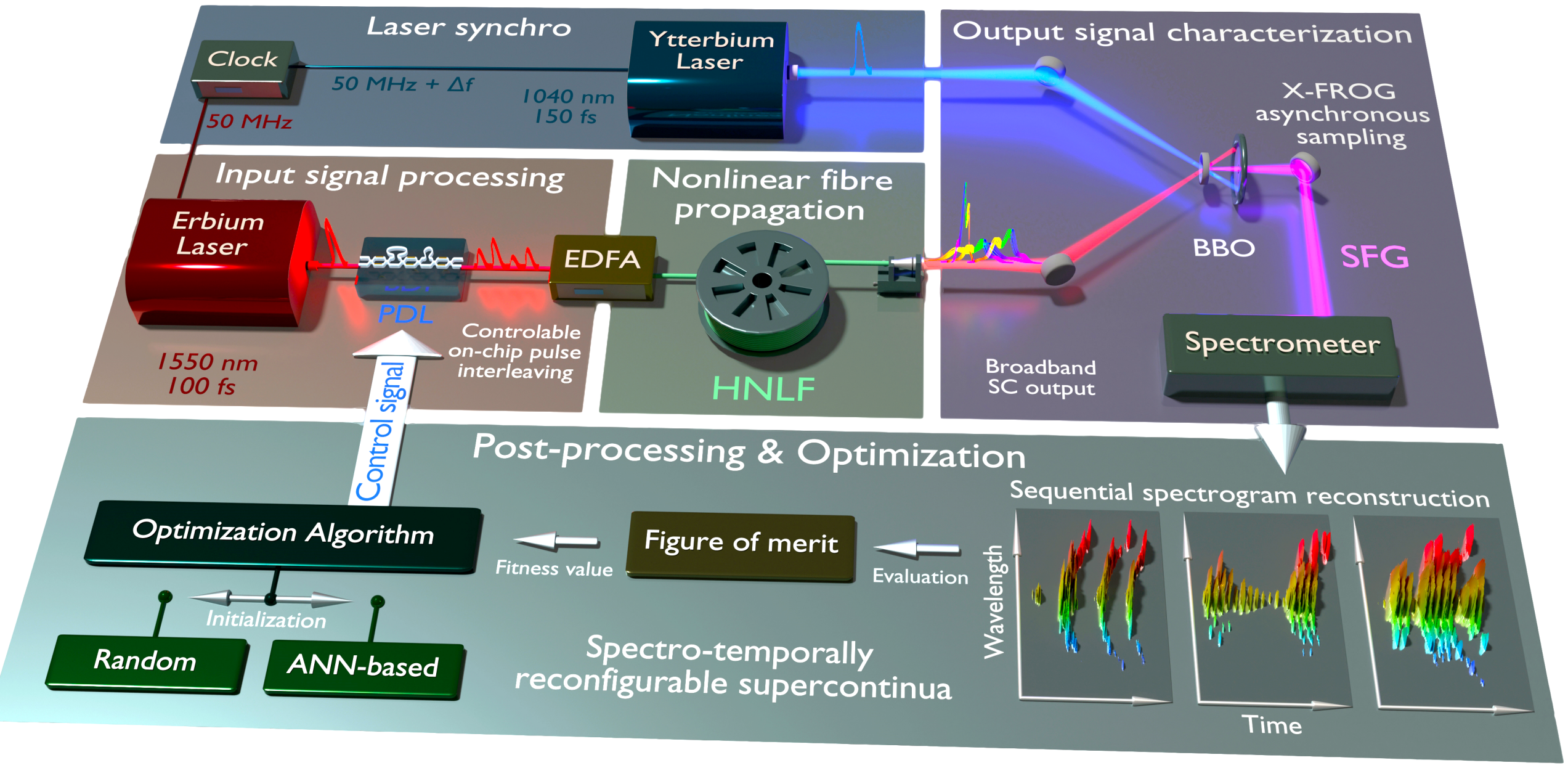

**Figure 2 | Experimental setup for tailoring broadband and reconfigurable wavepackets.** A signal processing stage first provides flexible control over a femtosecond input pulse train centred at 1550 nm through the use of an on-chip programmable delay line (PDL). The shaped pulse train is then amplified and launched into a highly nonlinear fibre (HNLF), where it undergoes nonlinear fibre propagation. The broadband wavepacket collected at the fibre output is a supercontinuum (SC) with controllable spectro-temporal properties. This broadband signal is then characterized via asynchronous nonlinear sampling, leveraging another synchronized ytterbium laser emitting at 1040 nm for producing sum-frequency generation (SFG). In such an X-FROG characterization stage, both lasers are frequency-mixed into a BBO crystal experimental spectrograms of the tailored broadband wavepackets. Three of such spectrograms are shown on the figure for illustrative purposes, resulting from three different settings of the input signal processing stage. Finally, in a post-processing and optimization stage, these spectrograms are rated with respect to a user-defined figure of merit (i.e. a selected spectro-temporal pattern), and then feed an optimization algorithm used to iteratively tailor the input pulse signal to reach a broadband wavepacket with the desired shape or properties.

The properties of this output wavepacket are monitored via a characterization stage (purple shading) relying on a cross-correlation frequency resolved optical gating (X-FROG)[50–53] setup, which is implemented via ASOPS: another femtosecond laser (at 1040 nm) is synchronized to the 1550 nm one, but exhibits a slightly different repetition rate (blue shading). This produces a temporal scanning effect between the two lasers that can be used, alongside a sum-frequency generation (SFG) process with a nonlinear crystal (BBO), for time gating. By sequentially recording the time-varying SFG signal with a fast spectrometer, we reconstruct the spectrogram of the output wavepacket (see Methods). The repetition rate detuning between the lasers is a key parameter, determining the trade-off between time resolution and acquisition speed. To allow for fast experimental optimization, we chose a measurement time of 10 s, which nonetheless allows for a sub-picosecond resolution of 922 fs.

This experimental approach exhibits an excellent variability in the spectro-temporal content that can be achieved in the various wavepackets generated. Yet, to truly take advantage of this flexibility towards practical applications, a method of autonomously searching the parameter space for a wavepacket with custom spectrogram profiles is necessary. To that end, we implement a post-processing and optimization stage (grey shadings in Fig. 2), consisting in the evaluation of a figure of merit (FoM) from the output spectrogram, and subsequent optimization of the initial pulse pattern (by iterative adjustment of the PDL settings). The FoM is based on a metric designed to concentrate power at a number of desired spectro-temporal locations (see Methods). We maximize it using a genetic algorithm (GA) and particle swarm optimization (PSO), with initial populations that are either randomly generated, or seeded by an ANN (see below)[54].

## Spectro-temporal optimization of optical wavepackets

To showcase the capabilities of our system for flexible wavepacket shaping, we summarize in Fig. 3 the results of five different spectro-temporal pattern optimization. The optimized spectrograms, measured experimentally over more than a 25 dB dynamic range and 400 nm spectral bandwidth, are displayed on the left panels, accompanied by their time and wavelength marginals. The target spectro-temporal positions (where the system aims to maximize intensity) are shown with colored circles, and as can be seen the optimized SC conform remarkably well to the desired patterns. The corresponding temporal profiles, at wavelengths of interest (defined for each specific FoM optimization), are shown on the right panels.

The results displayed in Fig. 3 correspond to SC waveforms obtained from GA optimization with random initialization (see Methods), and we stress that these solutions were found by the system in an autonomous fashion, with only the desired spectro-temporal positions being informed by the user. Through optimization, the system autonomously controls the nonlinear processes at play to place energy at the target locations (see Supplementary Media SM1).

For instance, in Fig. 3a,b, we target a pattern composed of two pulses at 1700 nm separated by 150 ps. This pattern, termed DT2, was chosen as a benchmark for its relative simplicity. Yet the interpulse delay, different from any predefined PDL delay (see Methods), already shows the flexibility of our system. Our system also allows us to control the number of pulses with a flexible temporal separation, as demonstrated in Fig. 3c,d: the pattern (DT4), aiming the generation of four pulses at 1700 nm, each separated by 60 ps, is successfully reached by GA optimization. Note that, unlike most other laser systems, here we can control the number of pulses without having to add beamsplitters or modify the optical setup.

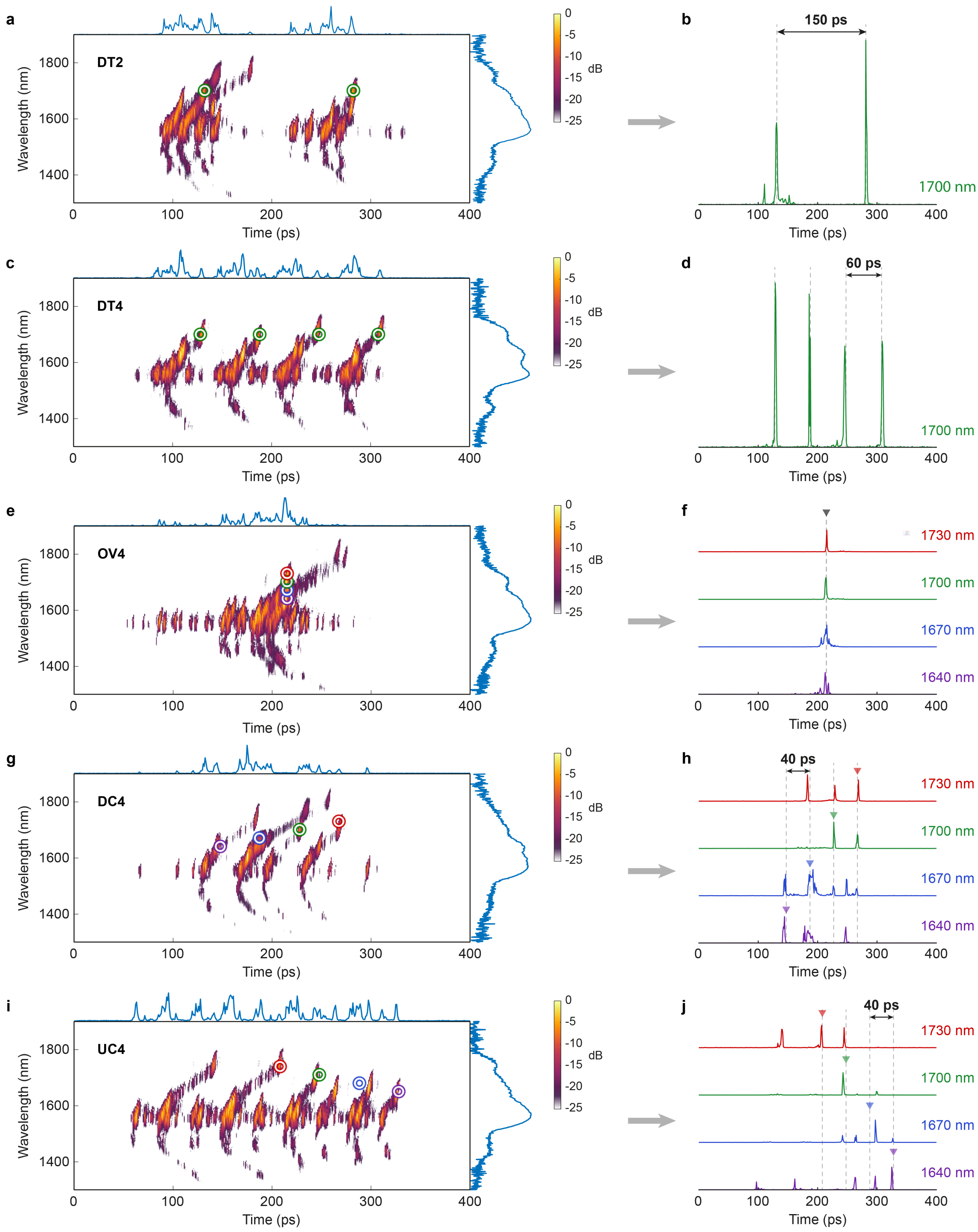

**Figure 3 | Examples of tailored broadband wavepackets, obtained via evolutionary algorithm optimizations.** Five different target spectro-temporal patterns are presented, to illustrate selected optimization of broadband wavepacket properties. Selective shaping and spectro-temporal power allocations are experimentally demonstrated, including the generation of selected number of pulses at desired wavelengths, as well as the control

of their relative wavelength detuning and delays. For each optimized wavepacket, the experimentally-measured spectrograms (in logarithmic scale) are shown on the left (along with the time and wavelength marginals at the top and on the right, respectively), and the desired pattern (i.e. positions of optimized energy) highlighted with colored circles in each spectrogram. The right panels show the corresponding spectrogram slices of the different wavelengths of interest for each pattern. Such temporal intensity profiles at selected wavelengths are here displayed in linear scale. The illustrated patterns respectively target: **a-b,** two pulses at 1700 nm separated by 150 ps (i.e. DT2); **c-d,** four pulses at 1700 nm separated by 60 ps (i.e. DT4) ; **e-f**, optimal temporal overlap of four pulses at different wavelengths (i.e. OV4); **g-h,** spectrally-encoded positive delay of 40 ps between four pulses at different wavelengths (i.e. DC4); **i-j,** spectrally-encoded negative delay of -40 ps between four pulses at different wavelengths (i.e. UC4).

In order to fully harness multiphotonic processes, highly sought-after for numerous applications, photons at different frequencies are required to synchronously interact. In Fig. 3e,f, with the pattern OV4, we show that the temporal overlap of four different wavelength components (1640 nm, 1670 nm, 1700 nm and 1730 nm) can be readily obtained through optimization. In this case, we experimentally demonstrate the joint excitation of multiple pulses, with broad (~100 nm) signal coverage, picosecond-level synchronization, and relatively large signal-to-noise ratio at the wavelengths of interest.

Pushing the complexity further, Fig. 3g-j shows examples of results where both temporal and spectral domains are controlled conjointly. Specifically, starting from the four wavelengths selected previously (within the pattern OV4), we experimentally show that we can implement a wavelength specific delay between different spectral components. Instead of aiming towards wavelength synchronization, here, each wavelength is respectively delayed from each other by 40 ps. In Fig. 3 g,h, this delay is positive, and increases with wavelength, in a manner analogous to a "down-chirp" pattern (i.e. DC4 – with a wavelength-specific delay beyond the dispersion profile of the fibre). Conversely, for the pattern displayed in Fig. 3 i,j, the delay is negative, which in this case corresponds to an "up-chirp" pattern (i.e. UC4 - with a wavelength-specific delay of sign opposite to the fibre dispersion). Due to the additional complexity of both the DC4 and UC4 patterns, here we see spurious pulses in addition to the desired profiles, but the target spectro-temporal pattern remains the dominant factor of the spectrogram. For completeness, more patterns, as well as PSO optimizations, are included in the Supplementary Material (Supplementary Note 1).

## ANN-informed optimization and speed-up

Notably, each optimization experiment displayed in Fig 3 took 3 hours, due to the necessity in measuring 1100 individuals (i.e. 1100 spectrograms). This requirement stems from the fact that we start with a random population, which might be far from the optimized individuals within the parameter space. To rectify this, we developed a hybrid workflow utilizing an ANN to kickstart the optimization process in a pre-informed way.

To train the ANN, we recorded a Monte Carlo (MC) dataset composed of 5,000 experimental spectrograms, acquired for random PDL splitting ratios. In this case, to speed up the acquisition and to assess the robustness of the proposed approach, we decrease the temporal resolution of the spectrograms by a factor 5, yielding a measurement of 2 s, for an equivalent resolution $\Delta\tau$ = 4.6 ps. The entire dataset is thus collected in just 3 hours, which corresponds to the duration of a single high-resolution pattern optimization presented in the previous section (Fig. 3). The ANN is a feed-forward network that takes PDL splitting ratios and outputs X-FROG spectrograms. Fig.

4 illustrates the ANN accuracy, comparing several synthetic spectrograms predicted from the ANN with their groundtruths (i.e. experiments) on a few examples drawn from the validation set.

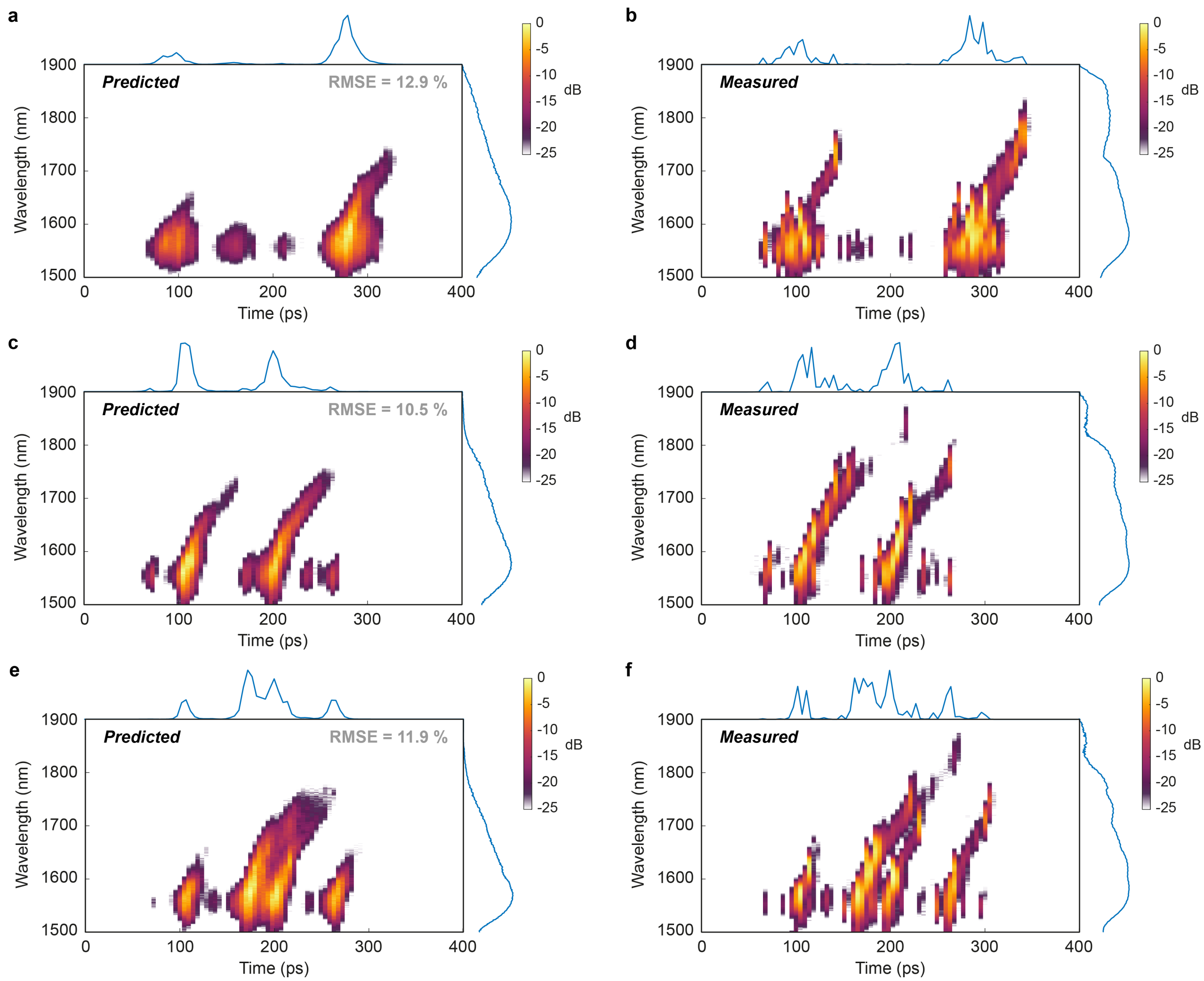

**Figure 4 | Artificial neural network prediction of synthetic spectrograms.** Synthetic spectrograms predicted from a trained artificial neural network (ANN) are compared with experimental X-FROGs (i.e. the ground truth data used for ANN training). The three examples illustrated are extracted from the ANN testing subset for different input pulse train configurations (left panels), and compared with the spectrograms measured experimentally (right panels). The spectrograms are shown on a logarithmic scale, with a 25 dB dynamic range, and an equivalent temporal resolution of 4.6 ps. The retrieved root-mean square error (RMSE), of typically few percent (12.2 % average RMSE in the testing subset), is displayed in each synthetic spectrogram.

These results demonstrate that the ANN successfully captures the general structure of X-FROG spectrograms. The overall agreement is fair (12.2 % ± 1.9 % RMSE prediction error over the whole testing subset), with the main features of the spectro-temporal shape being captured. However, sharp details, particularly in the long-wavelength edge, are often missing. We attribute this both to experimental drift and overall limited contributions from red wavelengths in the training set. The ANN is therefore not used as digital twin in our experiments, but rather as a tool to rapidly probe the parameter space. Due to its overwhelming speed when compared to the experimental bench, of 50 µs per spectrogram compared to 2 s in the laboratory, the ANN can quickly identify potential high-quality spectrograms for the target spectro-temporal profile, which allows the experimental optimization to be greatly sped up. The workflow of this hybrid approach is illustrated in Fig. 5a, which consists of two optimization stages: an offline numerical one and another online experimental one (see Methods). Typical optimization results (show for a DT4 spectro-temporal pattern) are displayed on Fig. 5b-d (see also Supplementary Media SM2).

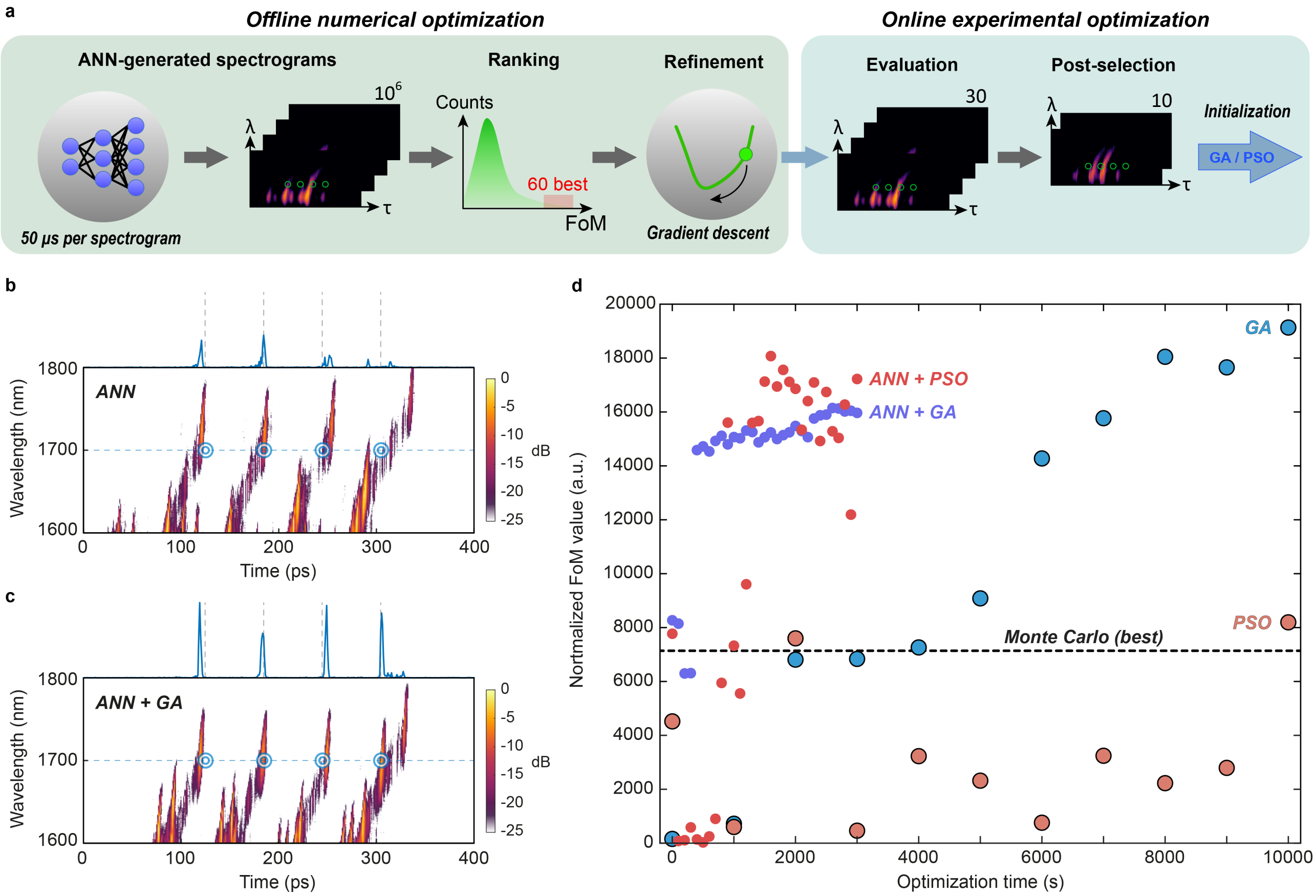

**Figure 5 | ANN-informed strategies for wavepacket optimization. a.** ANN-informed optimization workflow: once trained the ANN rapidly generates one million synthetic spectrograms. Each spectrogram is rated according to the FoM and the 60-best candidates are selected and further refined using fast numerical gradient descent optimization. The 30-best refined candidates are experimentally tested and the 10-best are selected as an optimized initial population for online optimization (via GA or PSO). **b-d.** Example of ANN-informed optimization of a selected target wavepacket pattern (DT4 – i.e. 4 pulses at 1700 nm, respectively delayed by 60 ps). **b.** Experimental spectrogram, measured from ANN-informed selection of optimal PDL settings. The spectrogram is part of the 10 best cases selected for DT4 target pattern, and used to feed the initial population towards online optimization. The target spectro-temporal pattern is shown with colored circles, and the temporal profile at wavelength of interest (1700 nm) is displayed on top. **c.** Corresponding experimental spectrogram, measured after online optimization. The spectrogram is obtained as a result of GA optimization from ANN-informed selection of initial population (as displayed in panel b). **d.** Evolution of FoM value over time, compared for different optimization techniques, for target DT4 spectro-temporal patterns. GA and PSO based on random initialization (circled disks) are compared to the same algorithms with ANN-informed initialization (dots). For clarity, the FoM values are normalized with respect to the median of a 5,000 Monte Carlo experimental dataset (i.e. $\text{FoM}_{\text{MC median}} = 1$). As a reference, we also provide the best case for DT4 pattern, within this MC dataset (dashed black horizontal line ; $\text{FoM}_{\text{MC best}} = 7{,}138$).

Fig. 5b shows the best post-selected spectrogram, measured experimentally from PDL settings predicted by the ANN (i.e. before any actual experimental optimization is performed). The corresponding spectrogram, after online optimization (i.e. GA with initial population provided by the ANN) is displayed on Fig. 5c. We likewise show the FoM evolution for different optimization processes in Fig. 5d, and further compare them to a MC dataset made of 5,000 high resolution ($\Delta\tau = 922$ fs) experimental spectrograms. Here the FoM values are normalized by the median of the MC data dataset, showcasing the large difference between optimized and typical values for the FoM.

We note that the various heuristic and hybrid approaches achieve a net gain of approximately twofold relative to the best MC result, despite the MC dataset requiring 50,000 s to be generated (just under 14 h). The hybrid ANN-based approaches further demonstrate a substantial speed-up, of approximately 15× in this case. This improvement can be understood from Figs. 5b–c, which compare the best ANN-selected individual before and after online optimization. The two spectrograms are already qualitatively similar, indicating that the ANN provides a high-quality starting point for the optimization. By initializing the search from such a favorable solution, the subsequent experimental optimization converges much more rapidly (see comparison of Supplementary Media SM1 and SM2).

This demonstrates the efficacy of ANN-seeded optimization, which generates a refined population at minimal computational cost, enabling rapid and reliable convergence with very few individuals. Once the ANN is trained, such hybrid technique thus allows the system to achieve user-defined spectro-temporal patterns in just minutes of system auto reconfiguration. Crucially, this approach remains effective even when performing wavepacket optimization on other spectro-temporal patterns, or at a finer temporal resolution than the ANN was initially trained on (see Supplementary Material – Figure S2). More importantly, the ANN used on Fig. 5 was trained on experimental data that were measured three weeks before the optimization experiment, showing both the stability of our system, and that eventual experimental system drift can be compensated by the quick and efficient online optimization.

Beyond individual case studies, we conducted a systematic statistical analysis of these hybrid optimization techniques across all spectro-temporal patterns, presented in Fig. 6. This figure shows the initial (shaded dots) versus optimized (triangles and stars) FoM values for all optimization techniques and patterns. We likewise include the histogram of the 5,000 MC dataset (grey violin plots) and its corresponding best value with dashed lines. This figure highlights the robustness of all the tested optimization techniques, which consistently achieve superior or comparable values relative to the maximum value from the MC dataset. From this figure, we likewise see that the initial populations of the hybrid techniques are clustered much closer to the optimized values, as opposed to randomly initialized optimization which largely follows the histogram of the MC dataset. This small FoM difference between initial and final optimized populations is the reason we observe a significant speed up in Fig. 5. We note that this hybrid workflow, exploiting an ANN (even if it acts as a coarse approximator) for experimental optimization, could be used in other scenarios when measurement time is a bottleneck but fast optimization is nevertheless desirable or required.

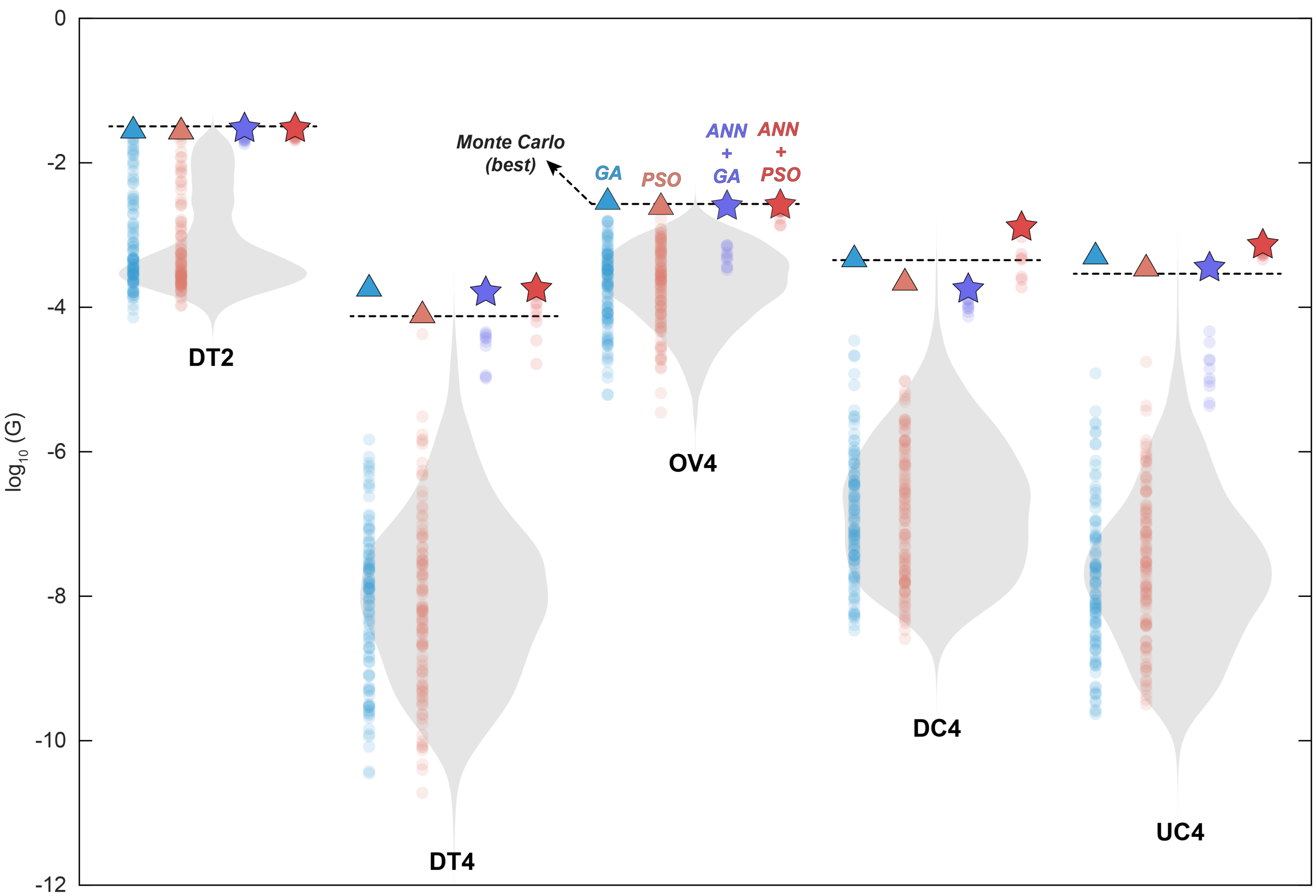

**Figure 6 | Benchmark of selected wavepacket pattern spectro-temporal shaping and optimization strategies.** Statistical analysis and comparison of FoM values achieved by ANN- and randomly-initialized GA and PSO optimizations across six spectro-temporal patterns with high temporal resolution (< 1ps). The x-axis represents the tested spectro-temporal patterns, as displayed in Fig. 3, while the y-axis shows the corresponding FoM values (i.e. G) displayed on a logarithm scale for clarity (note that, here, FoM values are not normalized). Light gray areas depict the probability density of FoM value, for each target pattern, computed from 5,000 spectrograms of an experimental Monte Carlo dataset, displayed in a violin plot format. Black dashed lines indicate the maximum values respectively retrieved from the best spectrogram within this MC dataset. Optimization methods are color-coded: cyan for randomly-initialized GA, orange for randomly-initialized PSO, violet for ANN-initialized GA, and red for ANN-initialized PSO. For each optimization, shaded dots represent the FoM values of initial population from the GA/PSO, and colored triangles/stars correspond to the final (optimized) FoM values.

## Discussion

Our system combines a programmable PIC with fast spectro-temporal characterization and ML-assisted optimization to control optical pulse trains that undergo nonlinear fibre propagation. By precisely and adaptively tailoring the temporal structure of the input pulse train, the nonlinear dynamics underlying spectral broadening can be actively controlled, enabling the generation of optical wavepackets with customizable spectro-temporal profiles. Experimentally, we demonstrate complex patterns including user-defined temporal delays at fixed wavelengths with an adjustable number of pulses, synchronized temporal overlap of multiple wavelengths, and wavelength-specific delays capable of overcoming the fibre's dispersion profile. The ability to exploit nonlinear broadening dynamics for on-demand control of broadband wavepackets over hundreds of nanometres and picosecond timescales represents a significant step towards programmable photonics.

The optimization workflow relies on metaheuristic algorithms operating directly on the controllable degrees of freedom of the programmable PIC. While conventional approaches rely on random population initialization, we introduce here an ANN-based initialization strategy that provides the optimizer with a high-quality starting population. The ANN enables rapid exploration of the parameter space, evaluating up to $10^6$ candidates in less than a minute (~50 μs per synthetic spectrogram), thereby shifting a substantial portion of the optimization process offline. This approach accelerates online experimental optimization by a factor of ~15, enabling the generation of tailored spectro-temporal profiles within minutes. Importantly, the approach remains effective across different target profiles and even in the presence of experimental drift: the ANN used for online optimization was trained on experimental data acquired three weeks earlier, demonstrating the robustness and temporal stability of the architecture.

Importantly, the framework does not inherently require the complex laser architectures or dual-stabilized oscillators used here for asynchronous X-FROG characterization. Instead, direct figure-of-merit measurements from the target application (for example, imaging signal-to-noise ratio or modality contrast - see Supplementary Note 3) could provide the optimization feedback, substantially broadening the range of potential applications, including biomedical diagnostics[55,56], coherent spectroscopy[10], and quantum technologies[43,44].

The platform also offers several avenues for further development. On-chip pulse-pattern reconfiguration could potentially reach kHz rates, given the experimentally measured MZI thermal settling time below 1 ms, while alternative tuning mechanisms, such as ultrafast thin-film lithium niobate switching[57], could enable substantially faster operation. For offline prediction, we employed a feed-forward neural network. Future implementations could benefit from more advanced ANN architectures or reinforcement-learning strategies. Furthermore, combining offline ANN pre-selection with online refinement could enable deployment on microcontroller units, allowing trained models and pre-optimized spectro-temporal patterns to be transferred from the laboratory to portable systems.

Although this work focuses on soliton-based interactions in the anomalous-dispersion regime, the demonstrated versatility suggests that other nonlinear dynamics could be exploited. Refined waveguide designs and tailored pulse characteristics could enable applications such as laser-cavity tuning and coherent frequency-comb generation[39]. Additionally, multimode waveguides[4] could provide an additional degree of freedom, namely the spatial one, to the controllable parameter space, eventually unlocking a comprehensive control of broadband light, benefitting applications such as advanced laser processing[40] and imaging[58].

Overall, the combination of programmable integrated photonics, ANN-assisted optimization, and nonlinear wave propagation provides a flexible platform for application-specific on-demand wavepacket engineering. Further advances in integrated waveguides, machine-learning architectures, and control speed could ultimately enable field-deployable systems for real-time, adaptive multidimensional wavepacket tailoring.

## Methods

**Experimental setup:** The setup, in general terms, consists of a mode-locked laser whose femtosecond pulses are launched into an on-chip programmable delay line (PDL, for reconfigurable pulse pattern generation), and is then amplified by an EDFA before propagation in a HNLF for nonlinear spectral broadening and supercontinuum (SC) generation. Specifically, the laser used is a 1550 nm fs laser (Menlo Systems - C-Fiber 780 HP) with a stabilized repetition rate of 50 MHz and an average power of 5.3 mW at the selected polarization-maintaining (PM) fibre port. The laser is subsequently coupled to 1 metre of PM Panda fibre (PM1550-XP), 1 m of standard single mode fibre (SMF-28) and 120 m of a dispersion compensation fibre (Thorlabs - DCF4). At 1550 nm, the DCF operates in the normal dispersion regime and is used to expand the duration of the pulse and minimize nonlinear effects before the HNLF. The chirped output of the DCF is then fed into an input port of the PDL. The output of the PDL (which consists of a train of pulses with customizable temporal pattern and a power level up to 1 mW) is then connected to an EDFA (Keopsys PEFA-SP) that amplifies the pulse train to an average power of 97 mW. The output of the EDFA is then connected to 8 m of SMF-28 to recompress the pulse into the HNLF and hence maximize peak power. This fibre dispersion management, with suitable choice of DCF and SMF-28 length throughout the setup, was designed to result in quasi transform-limited pulses of under 200 fs at the input of the HNLF, with mitigated nonlinear distortions during the initial pulse processing stage (see Supplementary Material - Supplementary Note 2 for detailed pulse characterization). The HNLF was fabricated in-house and operated in the anomalous dispersion regime (with a zero-dispersion wavelength around 1524 nm and a nonlinear coefficient $\gamma = 8.4$ $W^{-1}.km^{-1}$ at 1550 nm). The output of the HNLF was then connected to 3 m of SMF-28 fibre, up to a collimator which launched the broadband SC into the free-space X-FROG setup for spectro-temporal characterization (see details on X-FROG characterization below). The fibres, with the exception of the first metre right after the mode-locked laser, are not polarization maintaining.

**On-chip programmable delay line:** The on-chip PDL is fabricated on a CMOS-compatible, high-refractive index silica glass[59,60]. The chip is fully fibre-connectorized, with 6.7 dB overall optical coupling losses. It is formed of a cascade of 8 unbalanced waveguide sections, with each one splitting their input into two copies, and delaying one with respect to the other by a fixed amount. Each section has a delay that is double the delay of the previous section. The first unbalanced section has a delay of 1 ps and the last one of 128 ps, such that at the end, a train of up to 256 pulses spaced by 1 ps can be generated. Each of the unbalanced waveguide sections is preceded by a Mach-Zehnder interferometer (MZI), which controls the energy splitting ratio between the two subsequently delayed copies. The MZIs have two arms, which are both connected to electrodes that modulate the phase (optical path difference) of the MZI arms by a thermo-optic effect. The relative phase difference between the arms sets the MZI splitting ratios and, consequently, the temporal pattern of the output pulse train, such that this temporal optical pulse pattern be controlled electronically. Here, the electrodes utilize a push-pull architecture in a way that, at each MZI, the electronic power of one arm is matched by the other arm, so that the total electronic power dissipated by each MZI stays constant. Temperature exchanges on the PDL chip occur only locally (between respective MZI arms instead of between different MZIs), which greatly speeds up the thermal settling time (measured to be under 1 ms) of the system and further limit unwanted cross-talk for different PDL configurations. Due to this fast thermal settling time, through suitable microcontroller-based packaging, the PDL can be readily computer controlled via USB, and the optical pulse train can thus be reliably reconfigured at video-rate speed (>20 Hz). Like standard MZI-based electro-optic switches, MZIs are here characterized by two voltages values: $V_{bar}$, in which the optical pulses pass through the MZI without changing optical path (i.e. pulses coming from the long path at the $n^{th}$ unbalanced section are sent to the long path of the $(n+1)^{th}$ unbalanced section), and $V_{cross}$, in which optical pulses completely change optical path (i.e. pulses coming from the long path at the $n^{th}$ unbalanced section are sent to the short path of the $(n+1)^{th}$ unbalanced section). Through the use of optical taps at every unbalanced waveguide section, we characterized the transfer function and these two voltages for every MZI within the PDL. For our optimization experiments, we keep all MZI voltages on a range consisting of the interval between $V_{bar}$ and $V_{cross}$ plus an extra 20% margin (set at both sides of the transfer function). In other words, with $\Delta V = |V_{bar} - V_{cross}|$, the available MZI voltage settings , denoted V, will be in the interval $[V_{bar} - 0.2\ \Delta V\ ;\ V_{cross} + 0.2\ \Delta V]$ if $V_{cross} > V_{bar}$, or in the interval $[V_{cross} - 0.2\ \Delta V\ ;\ V_{bar} + 0.2\ \Delta V]$ if $V_{cross} < V_{bar}$.

**Spectro-temporal characterization:** The X-FROG setup is based upon using a second Ytterbium mode-locked laser emitting pulses with a temporal duration of 170 fs (FWHM) at a central wavelength of 1040 nm, with an average power that can reach up to 5 W (Menlo Systems - Orange Sync 50 MHz HP3). As illustrated in Fig. 2, both lasers are synchronized on a common GPS-disciplined oscillator (Stanford Research Systems - SRS FS752, with below -125 dBc/Hz phase noise), with repetition rates respectively tunable and actively-stabilized, ultimately leading to a temporal jitter below 90 fs between the femtosecond laser pulses of the two lasers. This custom, dual-oscillator and stabilized laser configuration is designed to perform asynchronous optical sampling[34] (ASOPS): the 1040 nm laser operates at a repetition rate slightly detuned from the repetition rate of the main (1550 nm) laser. This repetition rate detuning give rise to a temporal scanning effect, in which one laser is temporally- delayed with respect to the other at each laser period.

In our setup, asynchronous optical sampling is performed in a nonlinear fashion, analogous to a non-colinear intensity cross-correlation technique: both lasers are focused into a 100 µm-thick BBO crystal and a SFG signal is produced when both laser signals temporally overlap. The broadband output wavepacket around 1550 nm is sampled and mixed nonlinearly with the high-power femtosecond laser pulse at 1040 nm (with an average power up to 450 mW, adjusted experimentally to optimize the X-FROG dynamic range, acquisition speed and signal to noise ratio, while avoiding saturation during spectral detection). The SFG signal, generated around 622 nm, is then sent to a fast spectrometer (Ocean Optics - Ocean FX) to record SFG spectral content. In our setup, the time necessary to obtain a spectrogram (denoted $T_{scan}$) is given by

$$T_{scan} = \frac{1}{\Delta f}, \tag{1}$$

where $\Delta f$ is the repetition rate difference between the two lasers. Additionally, the temporal resolution $\Delta\tau$ of the spectrogram is given by

$$\Delta\tau = \frac{1}{\mathrm{R}\ T_{scan}} \frac{1}{\mathrm{f_{rep}}} = \frac{1}{\mathrm{R\ f_{rep}}} \Delta f, \tag{2}$$

where R and $\mathrm{f_{rep}}$ are respectively the spectrometer acquisition speed and the base repetition rate of the laser. In our experiment, those parameters are fixed, so that $\mathrm{f_{rep}} = 50$ MHz, and R $= 2170$ spectra/s (which corresponds to the spectrometer frame rate). One can see from Eq. (2) that there is a trade-off between measurement time and temporal resolution. A fine resolution necessitates a small detuning, while a fast measurement time requires a relatively high detuning (which can be experimentally adjusted between 500 μHz and 200 kHz). Since our experiments focus primarily on optimization, for most of the spectrograms of this work, we kept the measurement time at the manageable value of 10 s, which corresponds to a detuning of 0.1 Hz and a temporal resolution of 922 fs, which is enough to distinguish individual pulses produced by the PDL (with its minimum pulse temporal separation of 1 ps). The only spectrograms measured with a different temporal resolution were the ones used to train the ANN (see Fig. 4), in which case the measurement time was 2 s and the corresponding temporal resolution was 4.61 ps. Faster spectrogram measurements (potentially well-under a second) or finer resolutions (down to tens of fs, and mostly limited by laser relative jitter) could be readily achieved by simply tuning Δf, albeit with the expected trade-off between temporal resolution and speed.

We note that we do not perform a deconvolution (i.e. phase retrieval) on the measured X-FROG spectrograms, such as typically done in double-blind FROG[53]. This is because, here, we do not require the retrieved complex temporal profile of the output wavepacket, but rather a fast acquisition of the distribution of energy in the spectro-temporal domain, which is already provided by the spectrogram obtained with a 170 fs gate pulse (i.e. well-below the selected temporal resolution of the spectrogram). Another important observation is that, since our X-FROG system is based on SFG, its bandwidth is inherently limited by phase matching. However, due to the thin BBO crystal used, features can be observed over a bandwidth spanning well over 400 nm.

**Wavepacket spectro-temporal optimization target:** The figure of merit (FoM) used is based around concentrating power at specific spectro-temporal locations. These locations are specified with the vectors $\Lambda \in \mathbb{R}^p$ and $\mathrm{T} \in \mathbb{R}^{p-1}$, which respectively correspond to the spectral and temporal coordinates of the $p$ locations (i.e. peaks within the spectrogram) where one wishes to obtain high spectro-temporal intensities. The spectrogram is written $s(\lambda,\, t)$, where $\lambda$ is the wavelength coordinate, and $t$ is the time coordinate. The FoM then rewards spectrograms which have high power at locations given by vectors Λ and T. In this work, the function G used is the following:

$$\mathrm{G} = \mathrm{FoM}\,(\mathrm{a}(\lambda,\mathrm{t}),\Lambda,\mathrm{T}) = \frac{\max\limits_{\mathrm{t}}\left(\mathrm{a}\left(\Lambda_{\mathrm{p}},\,\mathrm{t}\right)\,\prod_{\mathrm{i}=1}^{\mathrm{p}-1}\mathrm{a}\left(\Lambda_{\mathrm{i}},\,\mathrm{t}-\mathrm{T}_{\mathrm{i}}\right)\right)}{\sqrt{\prod_{\mathrm{i}=1}^{\mathrm{p}}\int_{-\infty}^{+\infty}\mathrm{a}(\Lambda_{\mathrm{i}},\,\mathrm{t})^2\;\;\mathrm{dt}}}, \tag{3}$$

where $\mathrm{a}(\lambda,\mathrm{t})$ is a time-smoothed version of the experimental spectrogram $s(\lambda,\, t)$, and the terms $\Lambda_i$ and $\mathrm{T}_i$ are the vector elements of vectors Λ and T, respectively. We notice that since there is not an absolute time reference in the spectrogram, T is a vector of only $p-1$ elements, i.e. the last peak is taken as an implicit time reference. The time-smoothing to obtain $\mathrm{a}(\lambda,\mathrm{t})$ is performed by temporally convolving the spectrogram $\mathrm{s}(\lambda,\mathrm{t})$ with a gaussian window with a full-width at half-maximum (FWHM) of 20 ps. In other words, $\mathrm{a}(\lambda,\mathrm{t}) = \mathrm{s}(\lambda,\mathrm{t}) * \mathrm{e}^{-\left(\mathrm{t}^2/(2\sigma_{\mathrm{t}}^2)\right)}$ where $2\sqrt{2\,ln2}\,\sigma_t = 20$ ps. This approach was selected to make the figure of merit more tolerant, i.e. to give reasonably high FoM values to spectrograms that are a few pixels off the ideal pattern, which was observed to greatly improves optimization speed and robustness. The numerator of Eq. (3) is high when there are strong peaks at the desired locations, while the denominator is used to penalize spectrograms that feature a high numerator only because they possess significant energy at the wavelength of interests, and not because they correspond to the desired spectro-temporal shape. Finally, we stress that in the FoM formula of Eq. (3), only wavelengths in the vector Λ are taken into account, and wavelengths outside of this list do not directly contribute to the FoM evaluation. This is done because such wavelengths could, in principle, be selectively be filtered out of the optical wavepacket prior to the application at hand. For completeness, we also include below the vectors Λ and T for the spectro-temporal patterns shown in the main text.

| Pattern label | Number of desired peaks $p$ | Target wavelengths $\Lambda_i$ (nm) | Target relative delays $\mathrm{T}_i$ (ps) |
|---|---|---|---|
| DT2 | 2 | (1700, 1700) | (150) |
| DT4 | 4 | (1700, 1700, 1700, 1700) | (60, 120, 180) |
| OV4 | 4 | (1640, 1670, 1700, 1730) | (0, 0, 0) |
| DC4 | 4 | (1640, 1670, 1700, 1730) | (40, 80, 120) |
| UC4 | 4 | (1640, 1670, 1700, 1730) | (-40, -80, -120) |

**Metaheuristic search algorithms & optimization techniques:** To optimize the input PDL parameters for target X-FROG spectrograms using the previously defined FoM, two population-based metaheuristic algorithms were implemented: a Genetic Algorithm (GA) and Particle Swarm Optimization (PSO). Both algorithms operate in situ, directly interacting with the experimental system. Both algorithms also begin with an initial population, generated either randomly within the available parameter space, or informed by an artificial neural network (see below). At each or iteration, each candidate is evaluated by applying the PDL voltage to the setup, acquiring the corresponding X-FROG spectrogram, and computing its fitness with respect to the selected FoM.

Both algorithms are implemented in MATLAB using the Global Optimization Toolbox, with parameters set to the default toolbox values, and within the specified PDL voltage bounds. The GA evolves the population across generations using tournament selection, simulated binary crossover (SBX), and Gaussian mutation with an adaptive step size. Elitism is employed to preserve the best-performing individuals and prevent performance degradation across generations. Rank-based fitness scaling is used to maintain selection pressure throughout the generations. The elite count is set to 5 % of the population size, and the crossover fraction is 80 %. The PSO treats the population as a swarm of particles, each representing a candidate PDL vector. Particles update their positions and velocities based on both their personal best positions and the global best position discovered so far. A constriction factor and inertia weight scheduling are used to balance exploration and exploitation during the search process. The cognitive and social attraction coefficients are both set to a value of 1.49, and velocity clamping is automatically managed based on the bounds. The initial population of the optimization algorithms was either (i) random (with Monte Carlo generated PDL voltages taken from a uniform distribution), in which case the population size was 100 to ensure a good exploration of the search space, or (ii) from ANN-informed voltages (which serve as refined candidates for the spectro-temporal pattern at hand), in which case the initial population size was set to 10.

**Artificial Neural Network (ANN) training & hybrid initialization:** The mapping from PDL settings to XFROG spectrograms is learnt using a fully connected feed-forward neural network (FFNN). The network has 1 input layer, 3 hidden layers and 1 output layer. Each linear layer is followed by a LeakyReLU activation with a 1% negative slope, and a Dropout layer with a 5% probability to reduce overfitting (except for the output layer). The input layer receives the PDL voltage features (a 10-element vector). The three hidden layers contain 50, 500, and 5000 neurons, respectively. The output layer contains a number of neurons corresponding to the number of elements in each X-FROG spectrogram to predict (i.e. after suitable processing). Before training, the experimentally-recorded spectrograms are first cropped in wavelength and time, with a wavelength range between 1500 nm and 1900 nm, and a temporal window spanning 450 ps. After cropping, the total number of elements per processed X-FROG spectrogram is 34,986. Then, the spectrograms are converted to a logarithmic scale with a 30 dB dynamic range, and then normalized by the strongest pixel of the Monte Carlo dataset. These processed data are split into training (80%), validation (15%), and testing (5%) subsets. The ANN is then trained using a mean square error (MSE) loss and the Adam optimizer[61] with $\beta_1 = 0.9$ and $\beta_2 = 0.999$, along with an initial learning rate of $\eta_0 = 10^{-3}$. The learning rate is reduced upon plateau detection: by a factor of 2 if the relative validation loss is not reduced by 1% for 10 epochs, and with a 3 epochs cooldown. The ANN implementation and training workflow are built using PyTorch Lightning, enabling modular training loops, reproducibility, and streamlined GPU acceleration. The typical training time is approximately 6 minutes, for the 5000 data pairs and 150 epochs needed.

The trained ANN is used to rapidly generate one million synthetic X-FROG spectrograms from random PDL splitting ratios (50 µs per spectrogram on average, on a NVIDIA RTX 3090 GPU). Each spectrogram is evaluated on a parametrized FoM for the desired target spectrogram, and the 60 best-performing candidates per target spectrogram are retained in a dedicated buffer. These top 60 candidates for each target spectrogram then undergo a fast-numerical gradient descent refinement through the trained ANN to further refine their FoM. The Adam optimizer is used with $\beta_1 = 0.9$, $\beta_2 = 0.999$, and a fixed learning rate of $\eta_0 = 10^{-2}$, running for a maximum of 200 iterations, which takes approximately 1s per refined spectrogram (i.e. one minute to perform gradient descent refinement on the top 60 candidates selected by the ANN). The top 30 refined spectrograms are then converted back to their corresponding PDL splitting ratios and tested on the physical setup, where their experimental spectrograms are measured, and corresponding FoM values computed. Finally, the PDL splitting ratios leading to the top 10 experimental FoM serve as the initial population for ANN-initialized GA and PSO online optimization.

## Acknowledgments

This work received funding from the European Research Council (ERC) under the European Union's Horizon 2020 research and innovation programme under grant agreement No. 950618 (STREAMLINE project). B.W. acknowledges the support of the French ANR through the OPTIMAL project (ANR-20-CE30-0004) and the LabEx ∑-LIM (ANR-10-LABX-0074), and the Région Nouvelle Aquitaine (SPINAL project). DM acknowledges support from the Australian Research Council (ARC) Centre of Excellence Project in Optical Microcombs for Breakthrough Science (COMBS) under Grant CE230100006. The authors would like to sincerely thank Marc Fabert for experimental support, as well as Claire Carrion and Stéphanie Durand-Panteix for providing the biological sample shown in the supplementary material, and their assistance in its preparation. This research benefited from the support of the Platinom platform, with funding the European Union and the Nouvelle Aquitaine council under the PILIM program.

## Competing interests

The authors declare no competing interests.

# Spectro-temporal shaping of broadband optical wavepackets via programmable on-chip photonics

## *Supplementary Information*

**Bruno P. Chaves[1], Jérémy Saucourt[1], Van Thuy Hoang[1,2], Alexis Bougaud[1], Yassin Boussafa[1], Erwan Ferrandon[1], Sai T. Chu[3], Brent E. Little[4], David J. Moss[5], Roberto Morandotti[6], Vincent Couderc[1], Benjamin Wetzel[1*]**

*1. XLIM Research Institute, CNRS UMR 7252, Université de Limoges, Limoges 87060, France*
*2. School of Electrical and Mechanical Engineering, Adelaide University, Adelaide, SA 5005, Australia*
*3. Department of Physics, City University of Hong Kong, Tat Chee Avenue, Kowloon, Hong Kong, China*
*4. QXP Technologies Inc., Xi'an, China*
*5. Optical Sciences Centre, Swinburne University of Technology, Hawthorn, VIC 3122, Australia*
*6. Institut National de la Recherche Scientifique - Centre Énergie Matériaux Télécommunications, Varennes J3X 1S2, Canada*
**e-mail: benjamin.wetzel@xlim.fr*

### <u>Supplementary Note 1</u> - Versatile wavepacket optimization: flexible algorithm and adjustable temporal resolution

In the main manuscript, we discuss several optimization strategies and their application towards flexible spectro-temporal pattern generation. For completeness, we provide in Figure S1 and S2, examples illustrating the robustness of the proposed approach for different optimization settings.

In Figure S1, we first show that the results obtained by genetic algorithm (GA) optimization, as presented in Figure 3 of the manuscript, can also be obtained via a strategy based on particle swarm optimization (PSO). The spectrograms obtained in this case (using the same settings), show a qualitative agreement with the targeted features, and a good correspondence with the desired figure of merit (FoM) for all the configurations tested in Fig. 3 (see also Figure 6 for a quantitative comparison of the FoM values for each approach). We note that such a metaheuristic technique can also be initialized via synthetic dataset, generated by a trained artificial neural network (ANN) as discussed in the manuscript, so that to speed up the overall optimization process.

Figure S2 shows complementary wavepacket optimization results, for different FoMs that are not explicitly reported in the main manuscript. In this case, we use a randomly-initialized GA for optimization, but instead implemented a larger detuning between the two lasers within the asynchronous X-FROG characterization setup (with $\Delta f = 0.5$ Hz – see Methods). The corresponding spectrograms are thus measured with a temporal resolution of 4.61 ps, which drastically speeds up the measurement time (2 s per measured spectrogram). As observed in Fig. S2, the spectrogram resolution is indeed lower, but however sufficient to enable robust and efficient wavepacket optimization within a shorter timeframe.

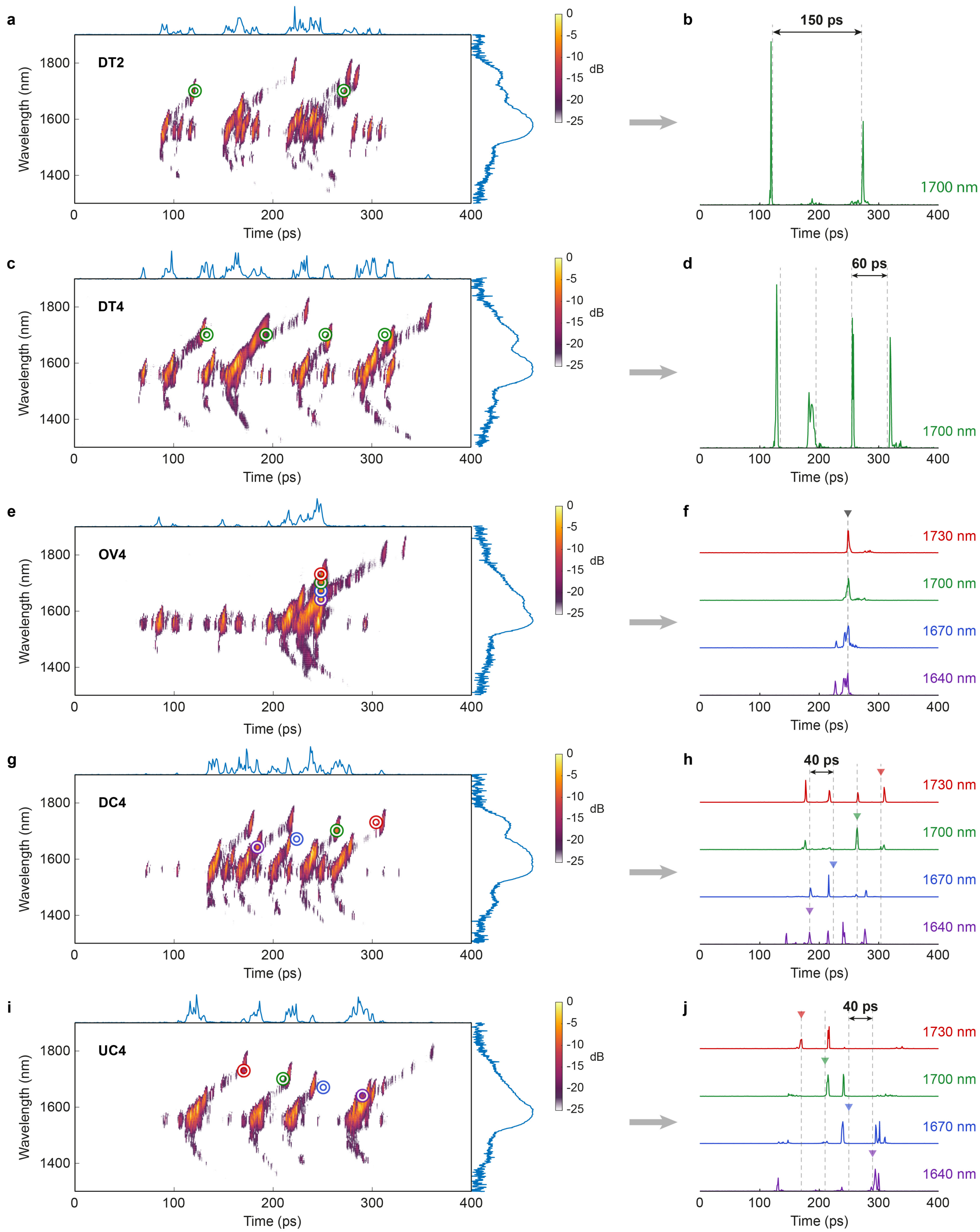

**Figure S1. Examples of tailored broadband wavepackets, obtained via particle swarm optimization.** The same five target spectro-temporal patterns are presented as reported in Figure 3 (within the manuscript), to illustrate selected optimization of broadband wavepacket properties obtained from randomly-initialized particle swarm optimization (PSO). For each optimized wavepacket, the experimentally-measured spectrograms (in logarithmic scale) are shown on the left (along with the time and wavelength marginals at the top and on the right, respectively), and the desired pattern (i.e. positions of optimized energy) highlighted with colored circles in each spectrogram. The right panels show the corresponding spectrogram slices of the different wavelengths of interest for each pattern. The temporal intensity profiles are here displayed in linear scale. The illustrated patterns respectively target: **a-b,** two pulses at 1700 nm separated by 150 ps (i.e. DT2); **c-d,** four pulses at 1700 nm separated by 60 ps (i.e. DT4) ; **e-f**, optimal temporal overlap of four pulses at different wavelengths (i.e. OV4); **g-h,** spectrally-encoded positive delay of 40 ps between four pulses at different wavelengths (i.e. DC4); **i-j,** spectrally-encoded negative delay of -40 ps between four pulses at different wavelengths (i.e. UC4).

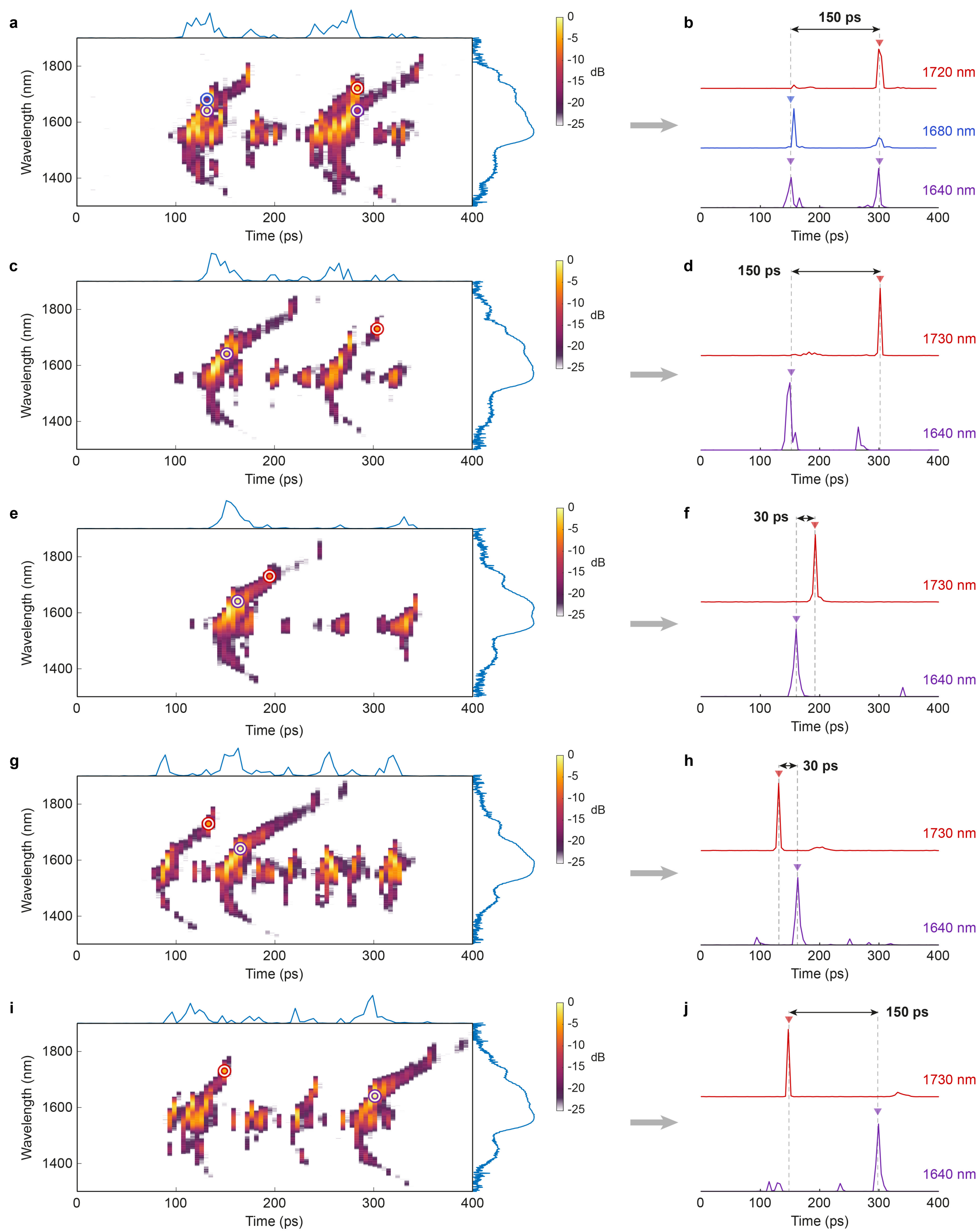

**Figure S2. Examples of tailored broadband wavepackets, obtained via genetic algorithm optimization but acquired with a lower temporal resolution.** For each optimized wavepacket, the experimentally-measured spectrograms (in logarithmic scale) are shown on the left (along with the time and wavelength marginals at the top and on the right, respectively), and the desired pattern (i.e. positions of optimized energy) highlighted with colored circles in each spectrogram. The right panels show the corresponding spectrogram slices of the different wavelengths of interest for each pattern. The temporal intensity profiles are here displayed in linear scale. The patterns illustrated here are complementary to those shown in the main text, and respectively target: **a-b,** temporal overlap of two pairs of different wavelength components, respectively separated by 150 ps from each other. In the other four patterns, we show the flexibility obtained in controlling the temporal separation between two wavelength components at 1730 nm and 1640 nm, respectively separated by: **c-d,** +150 ps; **e-f**, +30 ps; **g-h,** -30 ps; **i-j,** -150 ps.

The results displayed in Figure S2a,b illustrate that pairs of wavelength components (with different spectral detuning) can have an addressable picosecond delay, as typically required for successive synchronous multiphoton excitation with different energy level (see e.g. two-photon excitations or CARS excitation of different resonances displayed in Fig. 1 of the main manuscript). Conversely, the optimized patterns displayed in Figure S2c-j show that a delay tuning between isolated spectral components can be readily obtained and adjusted over several hundreds of picoseconds. This illustrate how this system can be leveraged towards addressable and sequential illuminations from different wavelength, without requiring mechanical delay lines or external synchronization systems. The demonstrated validity of such wavepacket optimizations opens the way towards refined optimization schemes, mixing different optimization schemes (ANN, metaheuristics, hybrid, etc.) with adaptive resolution. Ultimately, adjusting the asynchronous sampling 'on the fly' (and therefore the resolution and acquisition speed) should enable faster and even more convergence, in particular for applications requiring fast and versatile reconfigurability.

## Supplementary Note 2 - Dispersion management and input pulse properties

As described in Methods of the main manuscript, pulse dispersion management stage allows preparing suitable and reconfigurable femtosecond pulse(s) patterns before injection into the highly nonlinear fibre (HNLF), towards broadband supercontinuum generation. The mode-locked laser delivers pulses centred around 1550 nm at 50 MHz repetition rate (with 5.3 mW average power, and 106 pJ per pulse). The pulse propagates through 1 m of polarization-maintaining Panda fibre (PM1550-XP), 1 m of standard single mode fibre (SMF-28), and 120 m of dispersion compensating fibre (DCF - Thorlabs DCF4). At 1550 nm, the DCF operates in the normal dispersion regime and stretches the pulses to reduce their peak power, which limits nonlinear distortion during subsequent pulse processing and amplification. At the DCF output, the average power decreased to 3.8 mW, which corresponds to chirped picosecond pulses featuring 76 pJ energy. After 55 cm of SMF-28, the chirped pulses enter the on-chip programmable delay line (PDL). At the PDL output, the reconfigurable pulse pattern is amplified by an erbium-doped fibre amplifier (EDFA), and 8 m of SMF-28 then enable anomalous dispersion pulse recompression before the HNLF.

While the exact pulse properties (and associated power) depend on the pulse patterns generated by the PDL, the described dispersion management scheme ensure the recompression of each individual sub-pulse of the pattern into a ~150 fs pulse. For completeness, we report in Figure S3 the frequency-resolved optical gating (FROG) characterization performed at the HNLF input to ensure suitable pulse recompression. In this case, we monitored the configuration for which the PDL deliver a single pulse per laser period (i.e. acting similarly as simple optical waveguide element in a "all-pass" configuration – see Methods section). For the configuration of Fig. S3, the pulse average power is 0.8 mW (16 pJ) after the PDL. Following amplification and recompression, the average power at the HNLF input is 97 mW (1.94 nJ). Pulse measurements were performed and analysed using a commercial FROG system and deconvolution software (FEMTO Easy – MS-FROG) immediately before the HNLF. We emphasize that this input pulse measurement is completely distinct from the output X-FROG characterization of the supercontinuum reported in the manuscript, and further described in the Methods section. The experimental FROG trace is shown in Figure S3a, while the reconstructed trace following retrieval is presented in Fig. S3b (featuring a retrieval error of 0.067). The measured and reconstructed autocorrelation traces (Figure S3c) have full widths at half maximum (FWHM) of 209 fs and 226 fs, respectively, which corresponds to a relative difference of approximately 8.1% with respect to the measured value.

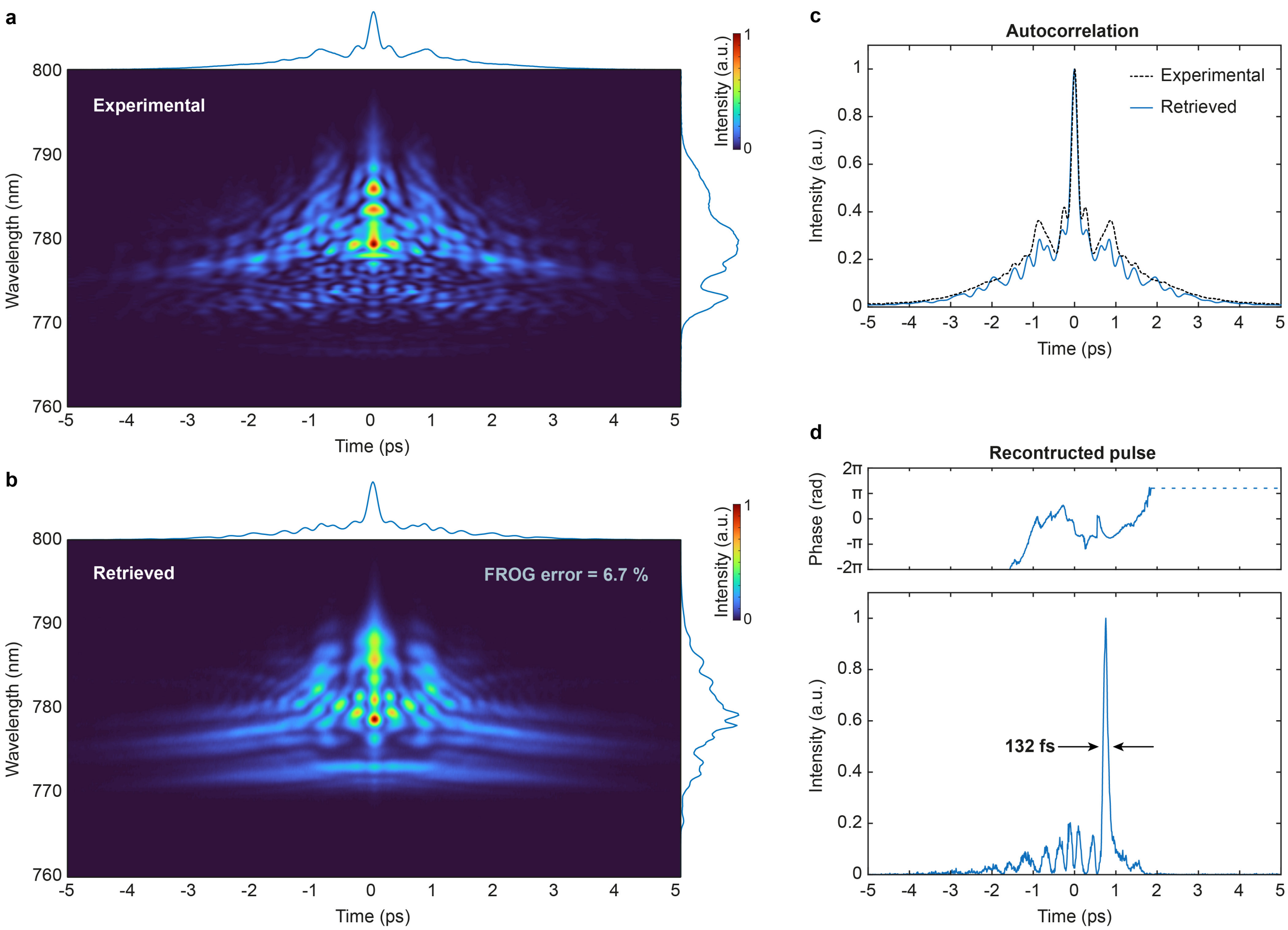

**Figure S3. Characterization of the dispersion-managed input pulse before injection in the HNLF. a,b**, Experimental (a) and retrieved (b) frequency-resolved optical gating (FROG) traces, obtained via second-harmonic-generation. The FROG traces are displayed on a normalized linear scale. The time and wavelength marginals, respectively corresponding to the autocorrelation traces and second harmonic spectra, are shown at the top and on the right of the traces. The FROG error after retrieval is 0.067. The measurement (performed at the HNLF input) corresponds to typical settings used in the manuscript for wavepacket optimization, but the on-chip programmable delay line (PDL) is here configured to deliver a single pulse per laser period (i.e. pass all configurations for all interferometers). **c**, Measured (dashed black line) and retrieved (solid blue line) intensity autocorrelations, with durations at full width at half maximum (FWHM) of 209 fs and 226 fs, respectively. **d**, Reconstructed pulse temporal phase (top) and intensity (bottom) obtained via commercial FROG pulse retrieval algorithm. The pulse main peak has a duration of 132 fs (FWHM), and is surrounded by a weaker pedestal and secondary temporal lobes.

The retrieved pulse temporal phase and intensity are shown in Figure S3d. The reconstructed pulse presents a predominant peak of 132 fs (FWHM), surrounded by a weaker pedestal (<20 % of the relative peak intensity) featuring small oscillations. This picosecond-long pedestal can be attributed to imperfect pulse recompression (in particular due to higher-order dispersion effects) as well as nonlinear distortions expected during EDFA amplification. The retrieved 132 fs intensity peak (and the agreement between measured and retrieved ~200 fs autocorrelations traces) provides a strong experimental reference for considering pulse duration below 200 fs to describe this single pulse recompression after dispersion management. This observation is also useful for gaining insight in the subsequent propagation dynamics (i.e. typical coherent supercontinuum generation associated with femtosecond pulse nonlinear propagation and spectral broadening). Furthermore, while other PDL configurations produce different temporal pulse patterns, the linear dispersive effect remains similar, so that sub-pulses are also following similar temporal chirp and recompressions, albeit minor changes from nonlinear distortions (arising from further energy spread within the overall picosecond-long pulse pattern).

## Supplementary Note 3 - Preliminary demonstration of reconfigurable multiphoton imaging

To illustrate the applicative potential of the programmable delay line (PDL), we performed a preliminary multiphoton imaging experiment in which a reconfigurable optical wavepacket was directly used for biological sample excitation. We note that these measurements are only provided here as an illustrative proof-of-principle, and should not be considered in light of the optimization results reported in the main manuscript. They nevertheless provide experimental indication that the reconfigurability enabled by the PDL can yield measurable changes in the response of a biological sample to versatile multiphoton excitation.

In the illustrative case reported below, the experimental architecture differs from the asynchronous X-FROG setup used throughout the manuscript. A similar dual-wavelength pumping scheme is used, arising from the same laser system (Menlo Systems - C-fiber and Orange HP). We however used different laser ports, and, more importantly, employed synchronous 50 MHz illumination, combining a fixed femtosecond excitation around 1040 nm with a second reconfigurable excitation centred around 1550 nm (with higher power and reduced spectral coverage compared to the settings of the manuscript). The 1040 nm laser delivers linearly polarized pulses with a duration of approximately 150 fs and an average power of 200 mW. The 1550 nm pulses (in the second branch) feature an initial average power of 200 mW and a duration of 150 fs at the laser output. This optical signal is first sent into a programmable spectral Fourier processor (Finisar – Wavesaper 4000A C+L) for dispersion and delay management, followed by the on-chip PDL for reconfigurable temporal processing. A subsequent erbium-doped fibre amplifier (EDFA) provides amplification (up to 300 mW) paired with moderate spectral broadening, so as to provide synchronized excitation covering the C+L wavelength region. Both excitation branches are then delivered to a laser-scanning multiphoton microscope (Evident - FVMPE-RS, with 25X objective Olympus - XLPLN25XWMP2) and the emitted signals are recorded in epi-detection using four photo-multiplier tubes (PMT) detectors in distinct spectral channels.

It is important to highlight that, unlike in the experimental setup discussed in the manuscript, here the 1550 nm laser pulse train only experiences mild spectral broadening (in the EDFA) and we do not have a fully-extended supercontinuum. Therefore, although the PDL does allow for great temporal manipulation, here the spectral control is quite limited, contrary to the system in the manuscript where the full spectro-temporal profile can be addressed. Nevertheless, the imaging setup here allows relatively high power and efficient multiphoton illuminations (mainly relying on the fixed femtosecond pump excitation around 1040 nm) while providing reconfigurability on the secondary (and synchronized) probe signal around 1550 nm. This synchronized configuration provides a simple way to observe multiphoton response without significant damages on the sample, at the expanse of not readily allowing X-FROG spectrogram measurement.

Figure S4 compares images of kidney (blood-vessel section) for two distinct PDL configurations. The four color-coded channels correspond to different and spectrally-detuned detection windows, which respectively span 410–465 nm, 495–540 nm, 575–640 nm and 750–800 nm.

The kidney sample is label free and enclosed into paraffin. These channels respectively contain contributions from multiphoton auto-fluorescence, second-harmonic generation (SHG) around 520 nm, and hybrid two-colour excitation with possible CARS-related responses. While the dominant sample morphology remains conserved between the two PDL settings (mainly provided by the 1040 nm excitation), localized differences are visible in several detection channels, particularly in the blue and near-infrared (red-color) responses. As a visual guide, several differences are highlighted with white-dashed circles in Figure S4e-h. These observations are consistent with a modification of PDL-induced excitation conditions with different temporal pulse patterns (and their associated spectral redistribution): images from PDL configuration #2 seems for instance to trigger a higher response (possibly including CARS responses) from the paraffin matrix and from specific regions surrounding the blood vessel.

This experiment illustrates a potential extension of the wavepacket optimization framework presented in the manuscript. Instead of evaluating each PDL configuration through the asynchronous X-FROG spectrogram, a signal measured directly from the biological sample could provide the required information towards practical light optimization for specific applications. Channel intensity or contrast, signal-to-noise ratio, or a combination of different imaging modalities could therefore be used as a specific figure of merit. In such a closed-loop implementation, the nonlinear response of the sample would effectively replace the X-FROG measurement as the optimization observable, enabling the programmable photonic system to search directly for illumination conditions that maximize a desired nonlinear response. While remaining beyond the scope of the present work, this constitutes a promising route towards dynamically reconfigurable microscopy or related applicative prospects.

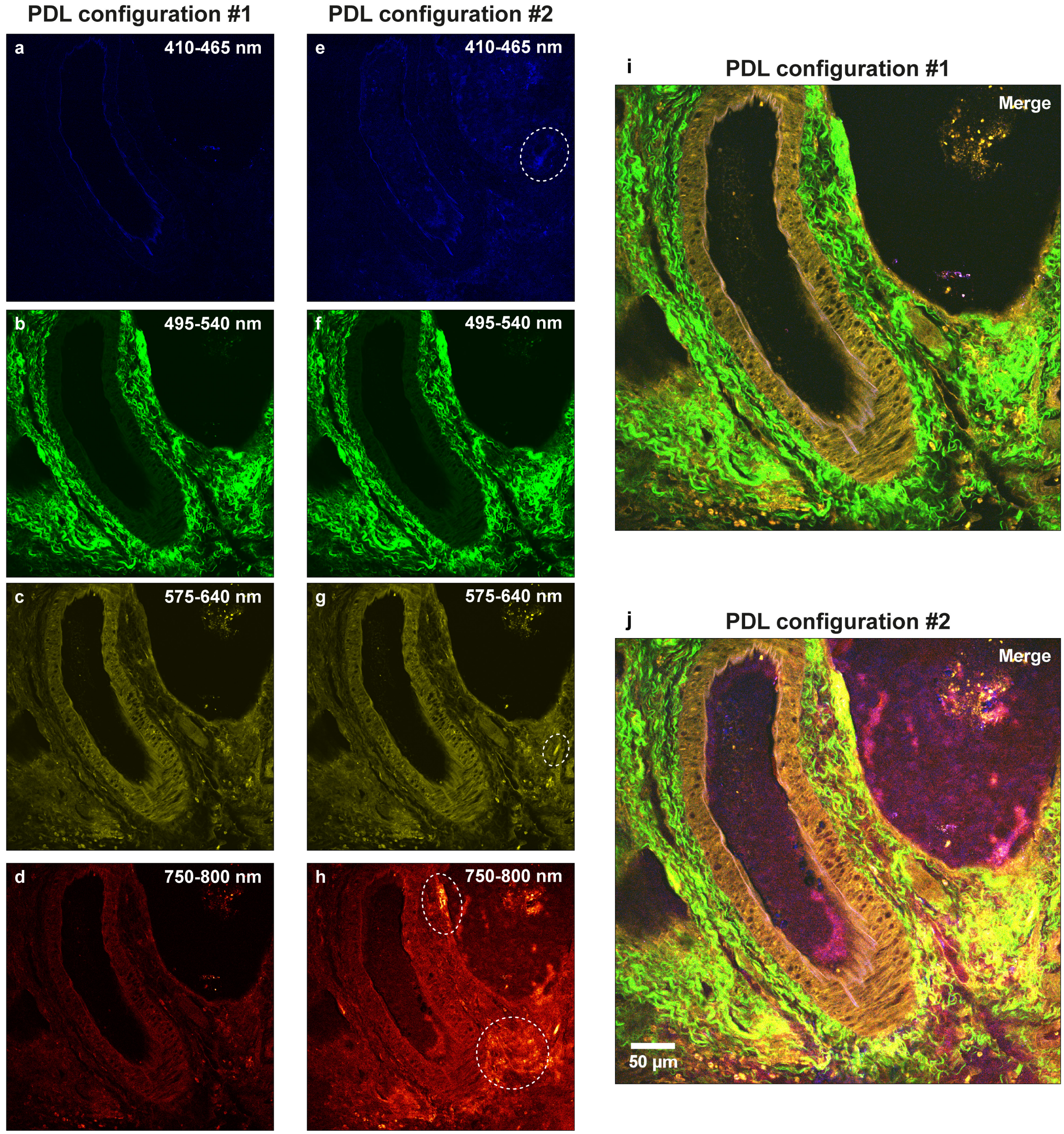

**Figure S4. Preliminary demonstration of reconfigurable multiphoton imaging using different PDL configurations.** Multiphoton microscope images of a label-free and paraffin-embedded kidney section, acquired under synchronized 1040 nm and 1550 nm excitation for two distinct settings of the on-chip programmable delay line (PDL). **a–d**, Images acquired with PDL configuration #1 in four epi-detection spectral channels spanning 410–465 nm (a), 495–540 nm (b), 575–640 nm (c), and 750–800 nm (d). **e–h**, Corresponding images acquired with PDL configuration #2 in the same detection channels. **i,j**, Composite images obtained by merging the four detection channels for PDL configurations #1 (i) and #2 (j), respectively. Each image was acquired with a resolution of 1024 × 1024 pixels and a 20 µs pixel dwell time. All acquisition parameters were kept identical between the two PDL configurations, and the corresponding images were processed and displayed using identical image-processing parameters, enabling direct qualitative comparison of the sample response. The different spectral detection windows contain contributions from multiphoton autofluorescence, second-harmonic generation (SHG), and hybrid two-colour multiphoton excitation (i.e. degenerate and non-degenerate two- and tree-photon absorption), with possible CARS-related contributions in the near-infrared detection channel. Several localized changes observed between the two PDL configurations are highlighted by white dashed circles. Scale bar of 50 µm is shown in (j).